\documentclass{aa}
\usepackage{graphicx}
\usepackage{txfonts}
\usepackage{psfig,longtable,lscape}
\usepackage{epsfig}
\usepackage{graphicx}
\usepackage{txfonts}
\usepackage[]{natbib}
\bibpunct{(}{)}{;}{a}{}{,} 
\def \xmm {{XMM-Newton}} 
\def \sax {{\em BeppoSAX}}

\def \glob {\mbox{NGC\thinspace6652}}
\def \src {\mbox{XB\thinspace1832--330}}

\def \hcm {\hbox {\ifmmode $ atom cm$^{-2}\else cm$^{-2}$\fi}}
\def \arcmin {\hbox{$^\prime$}}
\def \arcsec {\hbox{$^{\prime\prime}$}}
\def \chisq {$\chi ^{2}$}

\begin{document}
   \title{\xmm\ observations of the low--mass X--ray binary \src\ 
in the galactic globular cluster \glob}

   \author{L.\ Sidoli\inst{1}, N. La Palombara\inst{1}, T.\ Oosterbroek \inst{2}, 
A.N.\ Parmar \inst{3} }
   \offprints{L.\ Sidoli, sidoli@iasf-milano.inaf.it}
   \institute{INAF, Istituto di Astrofisica Spaziale e Fisica Cosmica, 
	Via E.\ Bassini 15,   I-20133 Milano,  Italy
    \and Science Payload and Advanced Concepts Office, ESA, ESTEC,
         Keplerlaan 1, NL-2200 AG, Noordwijk, The Netherlands
    \and Research and Scientific Support Department of ESA, ESTEC,
         keplerlaan 1, NL-2200 AG, Noordwijk, The Netherlands.
             }
   \date{Received 4 February 2008/Accepted 25 May 2008}
\abstract
{We report on two \xmm\ observations performed in 2006 of the luminous 
low--mass X--ray binary \src\ which is located in 
the galactic globular cluster \glob\ and is  probably
an ultracompact binary (orbital period, P$_{\rm orb}$, of $<$1 hour).
}
   {The aim of these observations is to investigate the low-energy absorption 
towards \src\ and in particular to 
search for Ne-rich material local to the binary, which has been suggested
as a possible spectral signature of  Neon rich degenerate companions.
}
{\xmm\ observed the source twice, in 2006 September and October. 
High resolution spectroscopy with the \xmm\ RGS was used to estimate 
the ratio of the neutral neon to oxygen column densities 
to search for an anomalous Ne abundance in this X--ray binary. 
}
{We find no evidence for anomalous Ne/O ratios, finding 
Ne/O=0.18$\pm{0.06}$ and  
Ne/O=0.17$\pm{0.03}$ (1$\sigma$ uncertainties),
in the two observations, respectively. These values  
are consistent with that in the interstellar medium. 
Timing analysis of EPIC data suggests possible periodicities at 9170$\pm{235}$~s and 
18616$\pm{531}$~s, which need confirmation.
A by-product of these observations consists of the spatial analysis of 
the source field, which resulted in the detection of 46 faint 
sources within the EPIC field of view, which are not present in the 
Second XMM-Newton Serendipitous Source Catalogue. 
All these faint sources are likely foreground objects.
}
{We performed the first high spectral resolution observations 
of \src, a probable ultracompact binary, without finding
any evidence for an anomalous Ne abundance. 
}
\keywords{X-rays: binaries - stars: neutron - accretion - X-rays: stars: individual: \src  }
\authorrunning {L.\ Sidoli et al.}
\titlerunning {\xmm\ observations of \src\ in \glob}

\maketitle

\section{Introduction}

\src\ was discovered using HEAO-1 \citep{Hertz1985} 
and is one of the bright low--mass X--ray binaries (LMXBs)
located in galactic globular clusters. 
The source was  observed with ROSAT (\citealt{Predehl1991}; 
\citealt{Johnston1996}) and associated with the 
galactic globular cluster \glob. 
The ROSAT uncertainty region was observed in 2000 May with 
the High Resolution Camera-Imager (HRC-I) camera  
on the $Chandra$ X-Ray Observatory \citep{Heinke2001}. 
Four X--ray sources were detected 
within 30$''$ of the cluster center, including the bright LMXB \src\ 
(named ``source A'' or CXOGLB~J183543.6-325926 in \citealt{Heinke2001}).
The detection of a type I X--ray burst confirmed 
the neutron star nature of the compact object  \citep{zand1998},
although one of the other fainter X--ray sources detected 
with $Chandra$ 
could also be responsible for the burst \citep{Heinke2001}.

The significantly smaller $Chandra$ uncertainty region allowed a secure identification of 
the optical counterpart \citep{Heinke2001} 
with a M${\rm _V}$ = +3.7 object (HST observations).  
A clear 0.08 mag variability in the V and I 
lightcurves was observed \citep{Heinke2001}, with possible periods 
of $\sim$3300~s, $\sim$8000~s (or $\sim$16000~s for ellipsoidal variations), as well
as non-periodic flickering. 
The short temporal coverature did not allow
the three periodicities to be clearly separated.
Heinke et al. also noted that the optical magnitude is  compatible with
the donor star being in an ultracompact binary 
(orbital period, P$_{\rm orb}$, less than about 1 hour).
If one of the three periodicity is associated with the orbital period,
the secondary could be a white dwarf or a low-mass M dwarf \citep{Heinke2001}.

X--ray spectroscopy of \src\ has been performed with ASCA \citep{Mukai2000} and
BeppoSAX \citep{Parmar2001} and indicates that no simple model provides
acceptable fits to the spectrum, 
suggesting that partial covering absorption, or other spectral
complexity plays an important role in modifying the X--ray emission at low-energies. 
Another more recent  observation at high energies (above 20~keV) has
been performed with IBIS on-board INTEGRAL \citep{Tarana2006}.

The galactic globular cluster \glob\ is located at a distance of 9.3~kpc, it 
has a metallicity of [Fe/H]=$-$0.9
and a low reddening ($E{\rm _{B-V} = 0.10 \pm 0.02}$) \citep{Ortolani1994},
which translates into an absorption,   
$N{\rm _H}$, of $\sim$$5.5 \times 10^{20}$~cm$^{-2}$
\citep{Predehl1995}.

We report here on the results of two \xmm\ observations performed in
order to investigate the low-energy absorption towards \src, in particular to 
search for Ne-rich material local to the binary, 
which has been suggested
as a possible spectral signature of  Neon rich degenerate companions 
in binaries  \citep{Juett2001}.

  \section{Observations and Data Reduction}

\src\ was observed with 
XMM-Newton on 2006 September 22 (hereafter, obs~I) 
and on 2006 October 20 (obs~II), for a net exposure of 
$\sim$28~ks and $\sim$34~ks, respectively (obs-IDs 0400790101 and 0400790301).
Data were reprocessed using version 7.1.0 of the Science Analysis
Software (SAS). Known hot, or flickering, pixels and electronic
noise were rejected. 
Response and ancillary matrix files
were generated using the SAS tasks {\em rmfgen} and {\em arfgen}.
The EPIC pn was operated in its Small Window mode, the MOS1 camera was in its Prime Partial Window mode,
with the central CCD operated in Timing Mode and with the other CCDs in Imaging Mode. The
MOS2 camera was operated in its Prime Partial Window with all the CCDs in Imaging Mode.
The EPIC MOS and pn observations used the medium thickness filter.
Spectra were extracted using PATTERN 0 for the MOS cameras 
(only single events), and PATTERNS from 0 to 4 
(single and double events) in the pn.
The RGS source and background events were extracted by making spatial and
energy selections on the event files.

Source counts were extracted from circular regions
of 40\arcsec\ radius for the pn and from a circular corona with an internal radius 
of 8\arcsec\ and an external radius of 1$'$ for MOS2. Source counts in
MOS1 timing mode were extracted from RAWX=309 to RAWX=330, excluding the central
core with an internal radius of 5 RAW pixels.
The MOS1 and MOS2 data clearly suffered from the effects of pile-up.
We checked iteratively with the SAS task {\em epatplot} that the events selected from 
annular regions 
in the MOS cameras were not significantly affected by pile-up after this spatial selection.
Background counts were obtained from similar sized regions offset from the source positions. 
For the MOS1 timing mode, the background region has been taken from an outer CCD 
which collected data in imaging mode.
The background did not show evidence for significant flaring activity.

To ensure applicability of the \chisq\ statistics, the
net spectra were rebinned such that at least 20 counts per
bin were present and such that the energy resolution was not
over-sampled by more than a factor 3.

  \section{Analysis and Results}

\subsection{Spectroscopy}

\begin{figure*}[th!]
\hbox{
\includegraphics[height=9.cm,angle=-90]{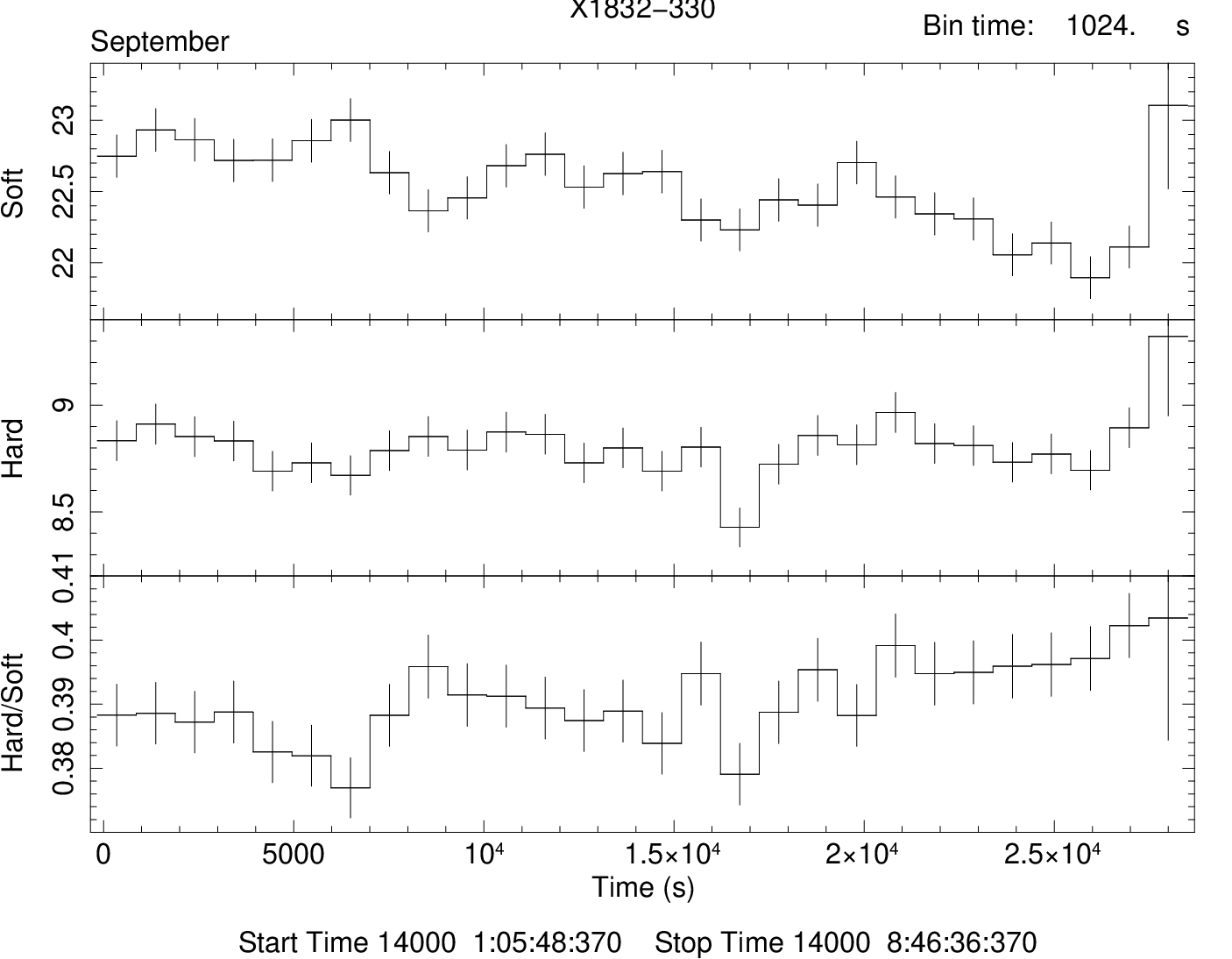}
\hspace{0.3cm}
\includegraphics[height=9.cm,angle=-90]{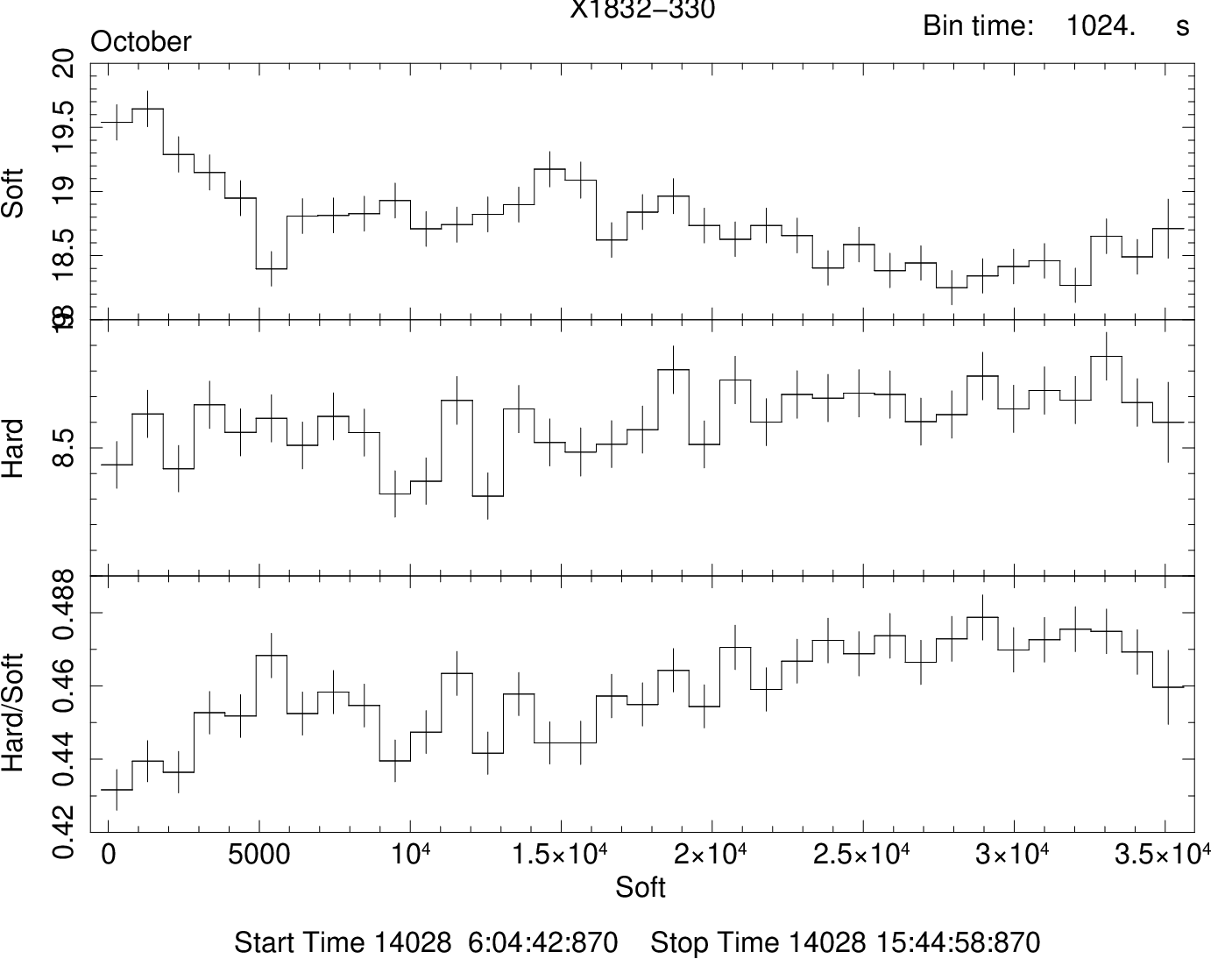}}
\caption[]{PN \src\ lightcurves of the two observations in the energy ranges 
0.3--2 keV and 2--12 keV. The bottom panels show the hardness ratios
(counts between 2--12 keV divided by those between 0.3--2 keV)
in the two observations. The binning is 1024~s. 
}
\label{fig:lc}
\end{figure*}

The \src\ lightcurves and hardness ratios do not show evidence for any large 
spectral variability (Fig.\ref{fig:lc}). We therefore
extracted averaged spectra from each of the two observations, which were then analysed separately,
withouth any further filtering or selection.

\begin{figure}[ht!]
\includegraphics[height=8.cm,angle=-90]{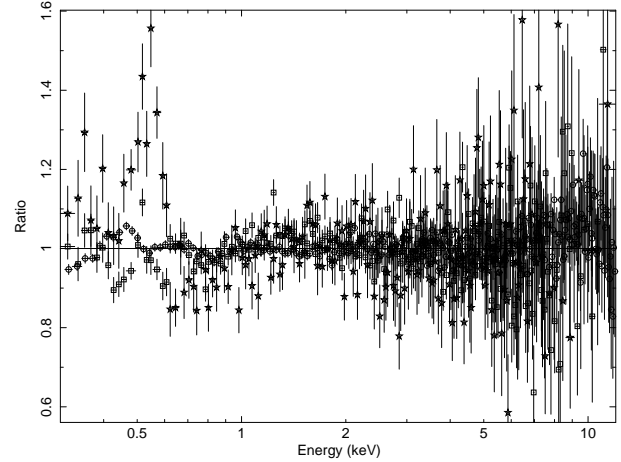}
\caption[]{Ratio between the data and the model when fitting
 the MOS1, MOS2 and pn spectra of obs~I with 
a double-component continuum (power law together with a blackbody) absorbed by a
partial covering fraction component, which accounts well for the global spectral shape of the
X--ray emission.
The different (and ``opposite'') residuals
in the pn (open circles), compared with the MOS1 (stars) and MOS2
(squares) are evident. 
A similar structured excess is present in the obs~II spectrum. 
}
\label{fig:ratio}
\end{figure}

Since the three EPIC instruments
show significant discrepancies below 0.8 keV when they are compared, 
with opposite structured residuals if fitted with the same spectral model (Fig.~\ref{fig:ratio}), 
we conservatively decided to limit the spectral analysis  
to the range 0.8--12 keV, for both the MOS and the pn.
Similar problems and discrepancies at low-energies have often been observed in other X--ray binaries
(e.g. \citealt{Sidoli2005x1850}; \citealt{Boirin2005}).
The softest energy range was covered with RGS1 and RGS2, which extend down to 0.4 keV.
For each observation separately, the two RGS (0.4--2 keV) spectra were fitted together with the
EPIC pn and the two MOS (0.8--12 keV) spectra. Free relative normalizations between the
instruments were included. 
The net source count rates are reported in Table~\ref{tab:log}.

\begin{table}[h!]
\caption[]{Net  count rates  (in units of s$^{-1}$) for  \src\ during the two observations.
The RGS rates are
in the energy range 0.4--2 keV while those from  EPIC are in the range 0.8--12 keV.
The exposure times are 28~ks (obs~I) and 34~ks (obs~II) for the EPIC MOS and RGS. 
The pn net exposures are $\sim$30\%  lower.
} 
\label{tab:log}
\begin{tabular}[c]{lcc}
\hline\noalign{\smallskip}
\hline\noalign{\smallskip}
                &   Obs~I &   Obs~II  \cr
\noalign{\smallskip\hrule\smallskip}
 RGS1            &    1.113 $\pm{0.007}$     &   0.874 $\pm{0.005}$    \cr
 RGS2            &    1.303 $\pm{0.007}$     &   1.064 $\pm{0.006}$    \cr
 pn                 &    32.38 $\pm{0.04}$      &   30.33 $\pm{0.04}$     \cr
 MOS1            &     4.42 $\pm{0.03}$      &   3.74 $\pm{0.02}$      \cr
 MOS2            &     4.23 $\pm{0.01}$      &   4.11 $\pm{0.01}$      \cr
 \noalign{\smallskip\hrule\smallskip}
\end{tabular}
\end{table}

Since the main aim of these \xmm\ observations was to investigate the low-
energy absorption towards \src, and in particular the estimate of the Ne/O ratio, 
we adopted a
variable abundance absorption model ({\sc vphabs} in {\sc xspec}),
with the elemental abundances set to the ISM values of \citet{Wilms2000}, 
except for those of O, Ne and Fe 
which were set to zero. Their effect on the spectral shape has been replaced
with three edges (O-K, Fe-L, Ne-K) with energies fixed at
0.543, 0.706 and 0.869 keV (e.g. \citealt{Paerels2001})
and edge depths that were allowed to vary.
In this way we could also account for a local Fe abundance which could be
different from the cosmic value. 

We tried different models for the continua of the two observations.
A single component model was never able to well fit the data adequately (as already
demonstrated by previous observations with ASCA \citep{Mukai2000} 
and BeppoSAX \citep{Parmar2001}).
Different combinations of simple models were tried:
for the soft component we adopted a blackbody and a disk blackbody 
model ({\sc diskbb} in {\sc xspec}),
while for the high-energy component we used a power law, a cutoff power-law and a 
Comptonization emission
model ({\sc compTT} in {\sc xspec}).
We tried all  possible combinations of these models, obtaining equally good fits and
similar spectral parameters.
Since in the \xmm\ energy band there is no evidence for a cut-off 
(resulting in an unconstrained electron
temperature in the {\sc compTT} model or in a high energy cut-off well above 10~keV), 
we adopted the power law plus a blackbody model for the continuum. 

From the depth of the edges and the elemental cross sections, the three element column densities 
were measured and so the Ne/O ratio, which has been compared to that of the
interstellar medium (ISM) value of 0.18, assuming the ISM abundances of \citet{Wilms2000}.
The spectral results are reported in Table~\ref{tab:spec} (for the continuum) and in 
Table~\ref{tab:edges} (for the edge
depths, columns and equivalent hydrogen columns, together with the final Ne/O ratio, for
the two observations).

\begin{table}[h!]
\caption[]{Spectroscopy of \src\  with the same model adopted
for the two observations (RGS1+RGS2+pn+MOS1+MOS2 data) consisting of
a power-law  model together with a blackbody, both equally absorbed by 
a variable absorption model, {\sc vphabs} in {\sc xspec}.
N$_{\rm H}$ is in units of 
10$^{22}$~cm$^{-2}$ and is the absorbing column density in {\sc vphabs},
with Ne, O and Fe abundances set to 0 (see Table~\ref{tab:edges} for the results on the neutral 
Ne, O, Fe edges). 
$\Gamma$ is the power law photon index,
kT$_{\rm bb}$ and  R$_{\rm bb}$ are the temperature (in keV) and the
radius of the blackbody emitter (in km 
assuming a distance of 9.3~kpc). 
Fluxes are in units of erg~cm$^{-2}$~s$^{-1}$ 
and are corrected for the absorption in the energy range 0.5--10 keV.  
L$_X$ is the 0.5--10 keV luminosity for the \glob\ distance.  Uncertainties are 
given at 90\% confidence.
} 
\label{tab:spec}
\begin{tabular}[c]{lcc}
\hline\noalign{\smallskip}
\hline\noalign{\smallskip}
Parameters                 &   Obs~I                &   Obs~II  \cr
\noalign{\smallskip\hrule\smallskip}
 N$_{\rm H}$               &    0.058 $\pm{0.006}$  &   0.094 $\pm{0.006}$  \cr
 kT$_{\rm bb}$             &    0.62 $\pm{0.01}$    &   0.61 $\pm{0.01}$    \cr
 R$_{\rm bb}$              &    3.1 $\pm{0.1}$      &   3.5 $\pm{0.1}$      \cr
 $\Gamma$                  &    1.58 $\pm{0.02}$    &   1.52 $\pm{0.02}$    \cr
 Flux                      &   1.6 $\times10^{-11}$ &   1.5$\times10^{-11}$ \cr
 L$_X$                     &   1.6 $\times10^{36}$  &   1.5 $\times10^{36}$  \cr
 red. $\chi^{2}$ (dof)     &   1.154 (2363)         &   1.044 (2712) \cr
\noalign{\smallskip\hrule\smallskip}
\end{tabular}
\end{table}

\begin{table}[!ht]
\caption[]{Results on the photoelectric absorption towards \src\ (see
Table~\ref{tab:spec} for the continuum emission parameters) when fitting RGS+pn+MOS spectra. 
The
edge energies were fixed at 0.54, 0.71 and 0.87~keV for O, Fe and
Ne, respectively. $\tau_{\rm edge}$ is the absorption depth.
$N_{\rm Z}$ is the element column density (in units of
10$^{17}$~\hcm) calculated using the Henke et al. \citep{Henke1993}
cross sections. $N_{\rm H}$ is the hydrogen column density implied
by $N_{\rm Z}$, in units of 10$^{21}$~\hcm, assuming the ISM
abundances of Wilms et al. \citep{Wilms2000}. Quoted uncertainties are at 1$\sigma$ confidence.}
\begin{center}
\begin{tabular}[c]{llllll}
\hline\noalign{\smallskip}
\hline\noalign{\smallskip}
Obs     & Edge & $\tau_{\rm edge}$           &  $N_{\rm Z}$            &  $N_{\rm H}$        & Ne/O \cr
\noalign{\smallskip\hrule\smallskip}
I       & Ne K & $0.040 ^{+0.009} _{-0.010}$ & $1.1 \pm{0.3}$          & $1.3 \pm{0.3}$      &                   \cr
        & O K  & $0.345 ^{+0.024} _{-0.029}$ & $6.1 ^{+0.4}_{-0.5}$    & $1.2 \pm{0.1}$      & $0.18 \pm{0.06}$  \cr
        & Fe L & $0.084 ^{+0.013} _{-0.014}$ & $0.12 \pm{0.02}$  & $0.45 \pm{0.07}$  &   \cr
\noalign{\smallskip\hrule\smallskip}
II      & Ne K & $0.055 ^{+0.009} _{-0.009}  $          & $1.5 \pm{0.3}$          & $1.7 \pm{0.3}$      &                   \cr
        & O K  & $0.505 ^{+0.025} _{-0.025}  $          & $8.9 \pm{0.4}$          & $1.8 {\bf \pm{0.1}}$      & $0.17 \pm{0.03}$  \cr
        & Fe L & $0.107 ^{+0.014} _{-0.014}  $          & $0.15 \pm{0.02}$        & $0.57 \pm{0.07}$    &                   \cr
\noalign{\smallskip\hrule\smallskip}
\end{tabular}
\end{center}
\label{tab:edges}
\end{table}

\begin{figure*}[th!]
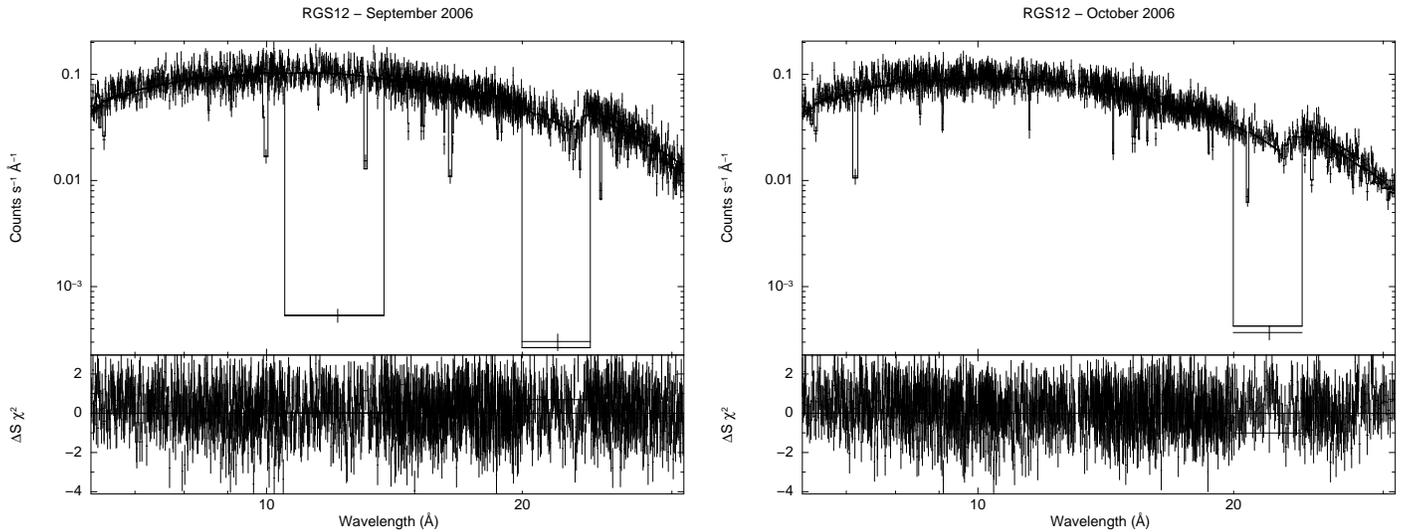

\hbox{
\includegraphics[height=9.cm,angle=-90]{fig3.ps}
\hspace{0.3cm}
\includegraphics[height=9.cm,angle=-90]{fig3b.ps}}
\caption[]{ RGS1 and RGS2 \src\ spectra in the two observations, fitted with the continuum model
used to fit the broad-band spectra (see text). 
The bottom panels show the residuals in units of standard deviations.
The missing regions in the RGS plots are due to the failed CCDs.
}
\label{fig:r12spec}
\end{figure*}

\begin{figure*}[th!]
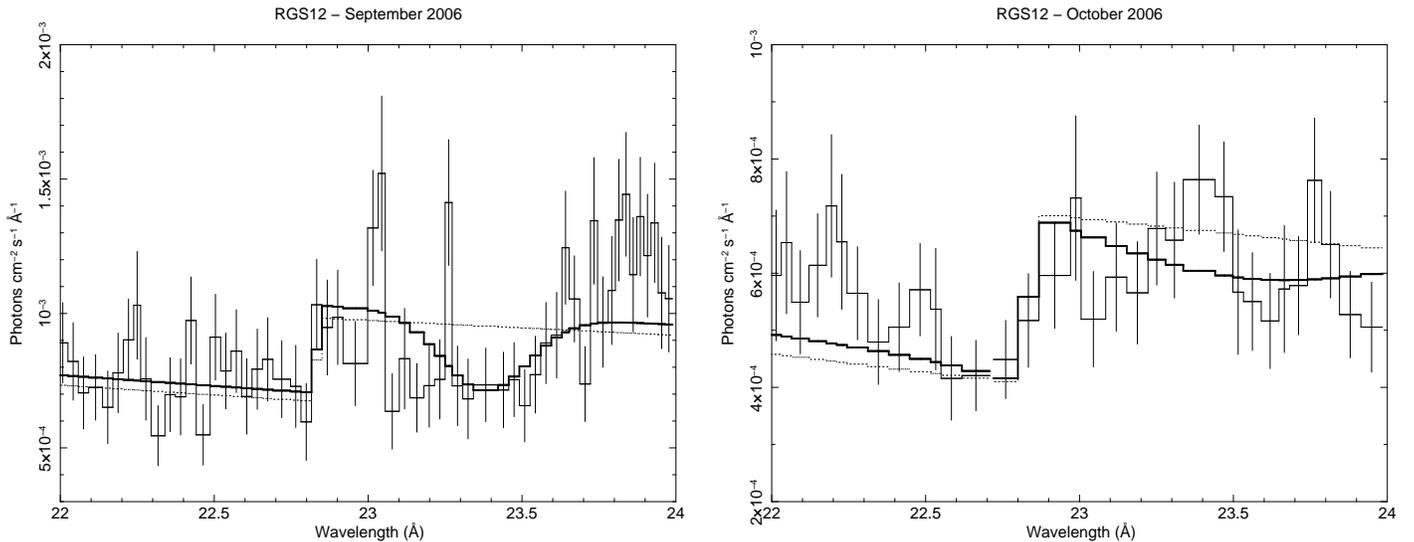

\hbox{
\includegraphics[height=9.cm,angle=-90]{fig4.ps}
\hspace{0.3cm}
\includegraphics[height=9.cm,angle=-90]{fig4b.ps}}
\caption[]{RGS1 and RGS2 \src\ spectra in the two observations, limited to the wavelenght region 
around the Oxygen edge. In the left panel, the excesses just above 23~\AA ~could be due to O~III emission line,
but it is less than 2 $\sigma$ from the continuum and the  statistics hampers a more detailed analysis.
}
\label{fig:r12specedge}
\end{figure*}

We then analysed the RGS spectra alone, to better investigate the fine structure 
around the absorption edges. We adopted the same shape for the continua
as found in the broader band spectroscopy, with temperature of the blackbody 
component and the power law
photon index fixed to the best-fits reported in Table~\ref{tab:spec}, 
but with all the normalizations free.
The same method to measure the edge depths of Ne, O and Fe was applied to the RGS1 and RGS2 spectra
(variable absorption model together with three edges). 
The best fit in both the observations (reduced $\chi$$^{2}$ = 1.105 (1735 dof) and 
$\chi$$^{2}$ = 0.937 (2070 dof), respectively) 
is obtained adding a further absorption feature, a Gaussian line
at energy of $\sim$0.53 keV, which is compatible with being due to  absorption from interstellar
O~I and O~II (see e.g. \citet{Juett2004}). 
In Fig.~\ref{fig:r12spec} and Fig.~\ref{fig:r12specedge} we 
show both the net RGS spectra and the
structure around the O edge. The absorption line at 0.53 keV has a normalization of
$-1.73(\pm{0.35})$$\times$10$^{-4}$~photons~cm$^{-2}$~s$^{-1}$ (Obs~I, EW=3$\pm{1}$~eV) 
and
$-1.51(^{+0.39} _{-0.41})$$\times$10$^{-4}$~photons~cm$^{-2}$~s$^{-1}$ (Obs~II, EW=5$\pm{1}$~eV).
The resulting depths for the absorption edges are the following:
$\tau_{\rm O }$=0.40$\pm{0.021}$,
$\tau_{\rm Ne }$=0.10 $^{+0.018} _{-0.020}$, 
$\tau_{\rm Fe }$=0.16 $^{+0.016} _{-0.017}$
in Obs~I, and 
$\tau_{\rm O }$=0.55$^{+0.029} _{-0.018}$,
$\tau_{\rm Ne }$=0.094$^{+0.022} _{-0.017}$, 
$\tau_{\rm Fe }$=0.185$^{+0.019} _{-0.016} $
in Obs~II (all errors on the depth of the edges are at 1~$\sigma$).
The resulting ratios Ne/O are 0.39 $^{+0.07} _{-0.08}$ (Obs.~I, 1~$\sigma$ uncertainty) and
0.27$^{+0.06} _{-0.05}$ (Obs.~II), which are within 2.6$\sigma$ and 1.8$\sigma$, respectively,
consistent with the ISM value of 0.18. Thus we 
confirm with high resolution RGS data alone, 
what found during the broader band analysis, with no evidence for an unusual ratio Ne/O.

\subsection{Timing analysis}

\begin{figure*}[th!]
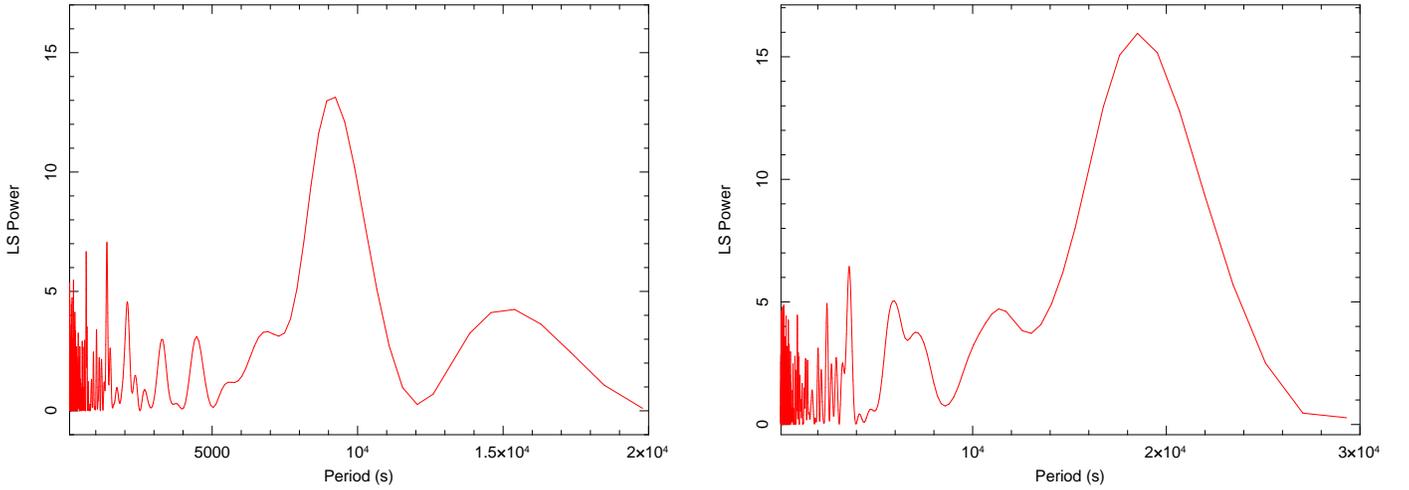

\hbox{
\includegraphics[height=9.cm,angle=-90]{fig5.ps}
\hspace{0.3cm}
\includegraphics[height=9.cm,angle=-90]{fig5b.ps}}
\caption[]{Lomb-Scargle periodograms of the two \xmm\ observations, 2006 September 
on the left and 2006 October on the right. 
}
\label{fig:timing}
\end{figure*}

\begin{figure}[ht]
\includegraphics[height=6.5cm,angle=0]{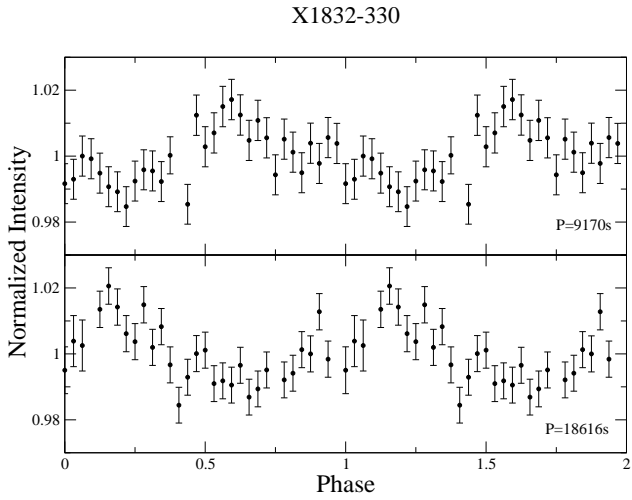}
\caption[]{Lightcurves of the two EPIC observations 
folded on the periods P = 9170~s (Obs~I, upper panel) and P = 18616~s (Obs~II, lower panel).
Phase zero is arbitrary. }
\label{fig:folded}
\end{figure}

From the Lomb-Scargle diagram of the obs~I 
there is evidence for a
periodicity near P = 9170$\pm{235}$~s, with a  significance of 
about 0.6~\% (see Fig.~\ref{fig:timing}, left panel). 
Folding on this period reveals a smooth sinusodial modulation with an amplitude
of 0.78$\pm{0.1}$~\%  (Fig.~\ref{fig:folded}).
This periodicity is not the
same as previously claimed values \citep{Heinke2001}, but periods in the 2--4
hour range have been claimed before. 
The period is rather long compared to the observation duration, 
thus we cannot say much about the long-term stability. 
However, it is quite likely (99.3~\%)
that during this observation an approximately sinusodial modulation with this period was present. 

In obs~II  there is no evidence 
for the P=9170~s periodicity. However, there is evidence
for modulation with a period of 18616$\pm{531}$~s. 
From the Lomb-Scargle periodogram
this period is significant at the 0.05~\% level (see Fig.~\ref{fig:timing}, right panel). 
It is striking that this period is twice the period found
in the first observation performed in 2006 September, within the uncertainties.

\subsection{Spatial analysis: a source catalog of the \src\ field}
\label{sect:det}

The EPIC images of the \src\ field show the presence of 
several faint X--ray sources. 
We therefore performed a detection analysis in order to obtain 
a source catalog of this region. 
We considered only the two MOS cameras, since the pn was operated in 
Small Window mode and could image a region close to the central target. 

The cleaned events files were used to produce MOS1 and MOS2 images 
in 5 energy ranges: 0.5--1~keV, 1--2~keV, 2--5~keV, 5--10~keV and 0.5--10~keV (total band). 
For each energy band, a corresponding set of exposure maps (i.e. one for each detector) was
generated with the SAS task {\em expmap}, to account for spatial 
quantum efficiency (QE), mirror vignetting and field of view variations. 
In order to maximize the {\em signal--to--noise} ratio ({\em S/N}) 
of our serendipitous sources and to reach lower flux limits, 
we merged the data (images and maps) from the two MOS cameras, for the two pointings separately. 
After the production of a detector mask (with the task {\em emask}), 
both images and exposure maps were used as a reference for the source detection, 
which was performed in four steps:

\begin{enumerate}
\item For each of the selected energy bands, the SAS task
{\em eboxdetect} was run in {\em local mode} to create a preliminary
source list. Sources were identified by applying the standard {\em
minimum detection likelihood} criterium, i.e., only candidate sources
with detection likelihood {\em -ln~P} $\ge$ 8, where {\em P} is the
probability of a spurious detection due to a Poissonian random
fluctuation of the background, were validated.

\item  Then, the task {\em esplinemap} was run to remove all the
validated sources from the original image and to create a background
map by fitting the so called {\em cheesed image} with a cubic spline.

\item For each of the selected energy bands the task {\em
eboxdetect} was run again, but in {\em map mode} using as a reference
the calculated background maps.

\item Lastly, the final source positions (with the EPIC combined MOS1+MOS2
count rates in each energy range) were calculated
using the task {\em emldetect}, which performs maximum likelihood
fits to the source spatial count distribution in all energy bands.
In this case, we checked that low threshold values of the equivalent 
single band detection likelihood (parameter (\textit{mlmin}) resulted 
in a high number of spurious detection. Therefore we fixed \textit{mlmin} = 30,
in order to select only meaningful sources.
\end{enumerate}

With the above procedure we found 25 sources in obs~I and 22 in the obs~II. 
In order to further increase the count statistics, we merged the data (images and maps) 
of the two pointings and applied the above detection procedure also to the merged data. 
In this case, using the same likelihood limit of \textit{mlmin} = 30, we detected a total of 41 sources.

\begin{figure}[ht]
\includegraphics[height=8.75cm,angle=0]{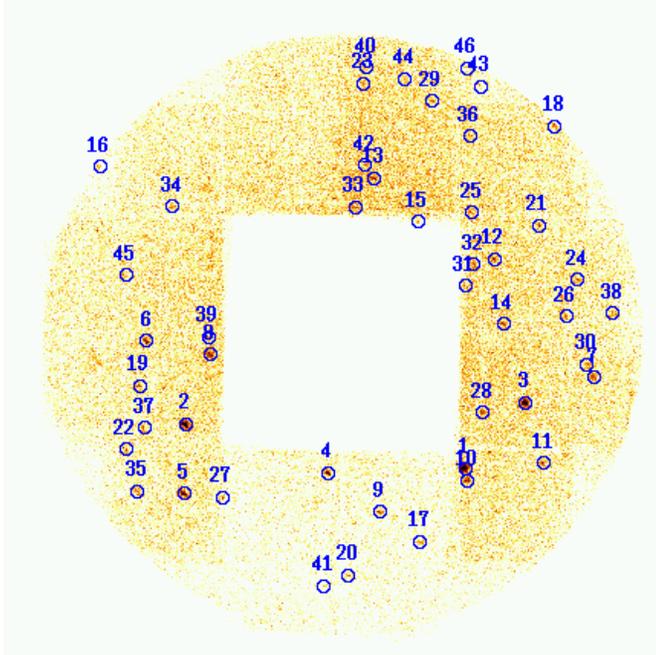}
\caption[]{Distribution of the 46 detected X--ray sources over the EPIC focal plane. 
The central CCD is empty since it was operated in Small Window mode and no source 
detection was performed on its data. Numbers mark the 46 sources listed 
in Table~\ref{sources}.} \label{source_distribution}
\end{figure}

Finally, we used the \textit{srcmatch} task to cross--match the three sets of detected 
sources to check for coincident ones. 
This list contains all the 46 sources, whose distribution over the observed sky 
area is shown in Fig.~\ref{source_distribution}. 
There we can check that sources \#15, \#16, \#18, \#31, and \#46 are close to 
the edge of the EPIC FOV, 
therefore we cannot exclude that they are spurious detections.

The main characteristics of these sources are summarized in Table~4. 
We list the source position (using the recommended XMM-Newton designations 
for serendipitous sources), the position uncertainty, the count--rate in the 
four fine energy ranges and the two hardness--ratios between, respectively, 
the two soft and the two hard ranges. 
In the reported position error we consider not only the statistical error, 
associated with the centroiding of the point source, 
but also the systematic error of 1'' \citep{Kirsch2007} 
due to the absolute positional accuracy of XMM-Newton. 
We also report the total source flux between 0.5 and 10 keV, which 
has been calculated from the total count rate in the full energy band. 
The applicable count--rate--to--flux conversion factor has been calculated 
assuming a power--law spectra with photon index, $\Gamma$, 
of 2 and the hydrogen column density in the direction of the globular cluster.

For each source we report if it was detected in the data of the 
single and/or the combined observations. 
In this way we find that 11 sources of the first 
observation are not detected in the second one, 
while 8 sources of the second observation are not detected in the first one. 
In order to verify if this result is due to a real source variability, 
we analysed both the position and count rates of these sources. 
We found that, in most cases, a source is detected in only one 
observation because in the other it either falls out of the imaged 
field--of--view (since the two observations are slightly rotated) 
or its count rate is just below the detection limit, even if the count rates 
measured in both the observations are consistent at 1 $\sigma$ confidence level. 
Only in few cases (highlighted with a 'V' in Table~\ref{sources}) 
we checked that the missing detection is due to a real source variability.
Regarding the sources detected in the merged observations, 
we find that only 14 (over a total of 41) are detected also in both 
the individual observations, while 8 and 6 are detected only in obs I and II, respectively; 
the remaining 13 sources are detected neither in the first 
nor in the second observation. 
Finally, we also note that 3 sources of obs I and 2 sources of 
the obs II are not detected in the merged observation.

We searched for possible known counterparts in the SIMBAD database, 
assuming an uncertainty radius equal to three times 
the source positional uncertainty. In this way we found no positionally coincident sources. 
The object known as NGC~6652~10, 
discovered by ROSAT \citep{Johnston1996}, is at a distance of 10.8$''$ from our source \#4. 
In Fig.~\ref{source_distribution} we note that this source is rather bright and, 
moreover, that there are no other X--ray sources near its position. 
Therefore, we suggest that it is indeed NGC~6652~10, whose position is now 
better constrained by the \xmm\ measurements. 

With the same search criteria, we searched  for possible counterparts in the 
GSC2.3 \citep{McLean+00}, USNO--NOMAD \citep{Monet+03} and 2MASS \citep{Cutri+03} catalogues. 
As shown in Table~4, with these constraints we found candidate 
counterparts for only 10 sources. In order to evaluate whether they are due to 
possible foreground contamination, we have to estimate the probability of chance 
coincidence between a X--ray and an optical source. 
This parameter is given by P = 1-e$^{-\pi r^{2} \mu}$, where $r$ is the X--ray 
uncertainty--circle radius and $\mu$ is the surface density of the 
optical sources \citep{Severgnini2004}. 
In our case, the surface density of the GSC2.3, USNO--NOMAD and 2MASS sources in the 
EPIC field--of--view is $~7.3\times10^{-3}$, $~1.7\times10^{-2}$, 
and $~3.6\times10^{-3}$ arcsec$^{-2}$, 
respectively. 
Based on the estimated position errors,  
we derived P $>$ 10 \% for all the candidate counterparts; 
therefore, we cannot exclude that they are due to random coincidences.

\section{Discussion}
\label{discussion}

We report here on two \xmm\ observations of \src\ (located in \glob) 
performed in 2006, about 1 month apart, which are the first high resolution spectra
of this globular cluster source.
\src\ is one of the luminous LMXBs (L$_{X}$$>$10$^{36}$~erg~s$^{-1}$)
located in a galactic globular cluster, and it is supposed 
to be an ultracompact binary, because of the
faintness of its optical counterpart \citep{Heinke2001}. 
Ultracompactness is also suggested by one of the possible periodicities
found in optical data \citep{Heinke2001}, if associated with the orbital period.

The 0.3--12 keV spectrum is complex and 
cannot be fit by a single model, requiring a soft 
component (here described by a blackbody with a temperature of 0.6 keV) 
together with  a hard power law (photon index of $\sim$1.5).
The spectral parameters of the soft component are similar to those
obtained by \citet{Parmar2001} during a \sax\ observation. 
The source displays also a similar 1--10 keV luminosity (with
\xmm\ we get $\sim$1.4$\times$10$^{36}$~erg~s$^{-1}$ in the same energy range).
The \xmm\ spectroscopy is compatible with what found by \citet{Sidoli2001globs}
in ultracompact binaries located in galactic globular clusters. 
Indeed, these sources have quite different properties 
from all the other bright globular cluster LMXBs with longer orbital periods.
This spectral analogy led \citealt{Parmar2001} to 
suggest that \src\ could be an ultracompact binary.

The total low-energy absorption resulting from the fits is compatible
with the
optically derived value in the direction of the host globular
cluster of $5.5\times$10$^{20}$~\hcm in the first observation, while
it is slightly higher in the second, suggesting an extra-absorption in the line of sight
of $\sim$3$\times$10$^{20}$ ~\hcm.
The likely presence of intrinsic neutral
extra-absorption  seems to be  confirmed by a good
fit  when using the partial covering fractional absorption model,
which indicates that \src\ is absorbed by an additional neutral
medium with a covering factor of $\sim$30\% and an intrinsic
hydrogen column density in the range 3--4$\times$10$^{21}$~\hcm\
for the two observations.

From the analysis of ASCA spectra of a few ultracompact binaries,
\citet{Juett2001} suggested the presence of 
an excess absorption of neutral Ne-rich material local to the
sources, leading some authors to propose that
the donor stars in some ultracompact binary systems
are Ne-rich white dwarfs (e.g., \citealt{Yungelson2002}). 
If this is the case, the high Ne/O ratio could be
a spectral signature of the presence of  some types of degenerate 
companions (e.g. Neon rich CO  white dwarf or 
O-Ne-Mg WD donors). 
High resolution spectra taken with \xmm\ and $Chandra$ 
confirmed a neutral Ne overabundance in a few cases
(e.g., 4U\,0614+091, \citet{Paerels2001}; 4U\,1543--624
and 2S\,0918--549, \citet{Juett2003}).
Instead, in other confirmed ultracompact sources, an
overabundance of neutral Ne was not found (e.g. 4U\,1850--087, \citealt{Sidoli2005x1850}
and \citealt{Juett2005}) with high resolution spectra, contrary
to the earlier ASCA measurement \citep{Juett2001}.
This  indicates a variability of the Ne/O ratio
probably due to luminosity variations (see e.g. \citet{Juett2003} or \citet{Sidoli2005x1850}). 
On the other hand, several known ultracompact binaries have never shown evidence
for unusual Ne/O ratios (e.g. 4U~1820--30, where a He white dwarf donor is present,
or 4U~0513--40, \citet{Juett2005}).

We adopted a variable absorption model together with three edges
corresponding to O-K, Ne-K and Fe-L features 
in order to measure the column density of the neutral
Ne, O and Fe in the line of sight. 
The high resolution \src\ spectra presented here do not show
any prominent emission features.
Our study of the strength of the edges of neutral Ne and O
reveals an Ne/O abundance ratio (see
Table~\ref{tab:edges}) which is consistent with the ISM value of 0.18 \citep{Wilms2000}. 
This implies that there is no evidence for the donor star in \src\ to be a
Ne-enriched white dwarf.

If we compare the measured elemental column densities  $N_{\rm Z}$ 
(converted to equivalent H column
densities using the Wilms et al.~(2000) ISM abundances) with the
$N{\rm _H}$ resulting from the overall shape of
the \xmm\ spectra there seems to be a
discrepancy, which 
could indicate an over-abundance of O and a sub-solar
abundance of Fe. This latter can be simply explained by the low
Fe abundance of the globular cluster. In any case we note that
we did not assume any uncertainty in the 
photoelectric cross-sections, which could be affected by 
an uncertainty as large as 30\% \citep{Paerels2001}. This could account for the observed
discrepancy. On the other hand, note that the resulting column density
is also dependent on the model adopted for the continuum. 
A Comptonization model, for example,
results into a lower absorption, because of the roll-over at soft energies. 

The timing analysis resulted into a possible periodicity of $\sim$2.5~hours 
in the EPIC lightcurve of the first observation, and into a second periodicity
(about twice that found in Obs~I) during Obs~II. The smallest period, if linked to the orbital
period of the binary system, would exclude an ultracompact binary (where orbital periods are less than about 1~hr). 
We note however that, although the period is significant, due to the limited duration of the
observations it is not possible 
to clearly say if this variability is a stable feature in the X--ray lightcurve,
or if it is a transient effect not linked to the orbital motion.

\begin{figure}[ht]
\includegraphics[height=8.75cm,angle=-90]{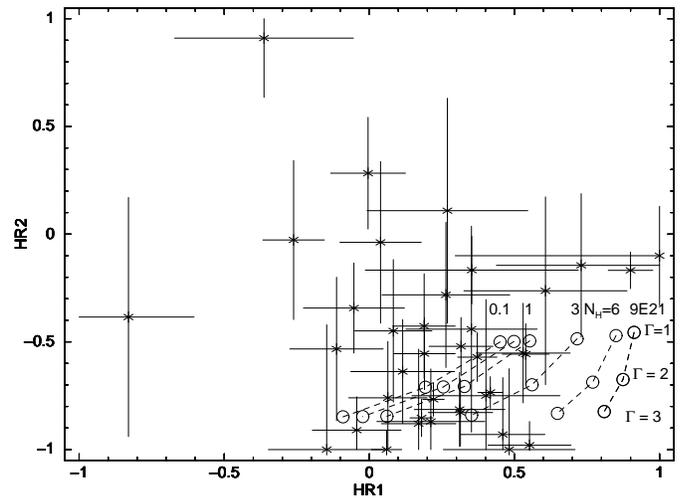}
\caption[]{Colour-colour diagram of the detected sources. 
Open circles mark the simulated colours 
(obtained with an absorbed power-law spectrum, assuming different 
photon indexes and absorption columns). 
From top to bottom the photon 
index is 1, 2, and 3; from left to right different column densities 
have been assumed: 0.1, 0.5, 1, 3, 6, 9$\times 10^{21}$ cm$^{-2}$}. 
\label{color-diagram}
\end{figure}

Our spatial analysis  has led to the compilation
of a catalogue of 46 faint X--ray sources in the \src\ field, most of which probably do not belong
to the globular cluster, as the \glob\ tidal radius is 4.48\arcmin\
(the 2003 version of the catalog is described in \citet{Harris1996}), 
well inside the central CCD of the MOS cameras, 
which was not used for source detection. 
These source are not
listed in the second XMM-Newton (2XMM) Serendipitous Source Catalogue (available online at
{\em http://xmmssc-www.star.le.ac.uk/Catalogue/2XMM}). 
In Fig.~\ref{color-diagram} we report the colour-colour diagram of 
the detected sources. For comparison, we overplot the simulated
colours obtained with an absorbed power-law spectrum, assuming 
different photon indices and absorption columns. 
Most of the sources appear to display a low absorption ($\sim 10^{20}$ cm$^{-2}$), 
lower than the  absorption towards \glob,
thus they are likely foreground objects.

\begin{acknowledgements}
Based on observations obtained with XMM-Newton, an ESA science
mission with instruments and contributions directly funded by ESA
member states and the USA (NASA).
This work was supported by  contract ASI/INAF I/023/05/0.
We thank the referee (A. Juett) for constructive comments and suggestions
which significantly improved our paper.
\end{acknowledgements}

\vspace{2cm}

\begin{landscape}

\begin{table}[htbp]
\begin{center}
\caption{The XMM--Newton Catalogue of sources in the region of X1832-330.}
\begin{small}
\begin{tabular}[c]{|c|c|c|c|c|c|c|c|c|c|c|c|c|}
\hline
(1)	& 	(2)			& 	(3)	& 	(4)			& 	(5)			& 	(6)			& 	(7)			& 	(8)			& 	(9)			& 	(10)			& 	(11)		& 	(12)	&	(13)	\\
SRC 	&	SOURCE NAME		&	ERR	& 	RATE 1			&	RATE 2			&	RATE 3			& 	RATE 4			& 	HR1			&	HR2			& 	FLUX			&	OBS		&	VAR	&	CTP	\\
ID	&				&	($''$)	& ($10^{-3}$c s$^{-1}$)		& ($10^{-3}$c s$^{-1}$)		& ($10^{-3}$c s$^{-1}$)		& ($10^{-3}$c s$^{-1}$)		& 				& 				& 	($10^{-14}$ cgs)	&			&		&		\\ \hline
1	&	XMMU J183516.1-330602	&	1.02	&	2.85	$\pm$	0.18	&	4.42	$\pm$	0.21	&	1.77	$\pm$	0.14	&	0.22	$\pm$	0.09	&	0.22	$\pm$	0.04	&	-0.77	$\pm$	0.08	&	11.98	$\pm$	0.56	&	I,II,I+II	&	/	&	0;0;0	\\
2	&	XMMU J183619.4-330357	&	1.03	&	2.46	$\pm$	0.17	&	3.37	$\pm$	0.19	&	1.29	$\pm$	0.13	&	0.15	$\pm$	0.09	&	0.18	$\pm$	0.04	&	-0.85	$\pm$	0.09	&	10.45	$\pm$	0.53	&	I,II,I+II	&	/	&	1;0;1	\\
3	&	XMMU J183502.4-330257	&	1.03	&	1.47	$\pm$	0.13	&	3.52	$\pm$	0.19	&	2.09	$\pm$	0.15	&	0.31	$\pm$	0.10	&	0.42	$\pm$	0.04	&	-0.73	$\pm$	0.07	&	8.99	$\pm$	0.50	&	I,II,I+II	&	/	&	0;0;0	\\
4	&	XMMU J183546.9-330617	&	1.06	&	1.22	$\pm$	0.16	&	2.63	$\pm$	0.22	&	1.36	$\pm$	0.17	&	0.38	$\pm$	0.12	&	0.37	$\pm$	0.07	&	-0.57	$\pm$	0.11	&	7.23	$\pm$	0.57	&	I,II,I+II	&	/	&	1;0;0	\\
5	&	XMMU J183619.7-330714	&	1.05	&	2.37	$\pm$	0.18	&	2.66	$\pm$	0.19	&	0.38	$\pm$	0.09	& \textit{0.00	$\pm$	0.04}	&	0.06	$\pm$	0.05	&	-1.00$^{+0.09}_{-0.00}$	&	7.22	$\pm$	0.53	&	I,II,I+II	&	/	&	0;0;0	\\
6	&	XMMU J183628.0-330001	&	1.10	&	0.07	$\pm$	0.06	&	1.26	$\pm$	0.14	&	1.54	$\pm$	0.15	&	1.14	$\pm$	0.17	&	0.90	$\pm$	0.08	&	-0.17	$\pm$	0.08	&	3.97	$\pm$	0.40	&	I,II,I+II	&	/	&	1;0;0	\\
7	&	XMMU J183446.9-330143	&	1.19	&	0.55	$\pm$	0.12	&	1.00	$\pm$	0.13	&	0.93	$\pm$	0.13	& \textit{0.06	$\pm$	0.08}	&	0.31	$\pm$	0.11	&	-0.83	$\pm$	0.16	&	3.25	$\pm$	0.41	&	I,II,I+II	&	/	&	0;2;0	\\
8	&	XMMU J183613.6-330039	&	1.11	&	0.38	$\pm$	0.08	&	1.27	$\pm$	0.11	&	0.68	$\pm$	0.09	&	0.19	$\pm$	0.07	&	0.54	$\pm$	0.08	&	-0.56	$\pm$	0.14	&	2.93	$\pm$	0.30	&	I,II,I+II	&	/	&	0;0;0	\\
9	&	XMMU J183535.3-330808	&	1.14	&	0.84	$\pm$	0.15	&	1.24	$\pm$	0.17	&	0.93	$\pm$	0.15	&	0.23	$\pm$	0.12	&	0.19	$\pm$	0.11	&	-0.56	$\pm$	0.17	&	4.05	$\pm$	0.49	&	I,II,I+II	&	/	&	0;2;0	\\
10	&	XMMU J183515.5-330637	&	1.12	&	0.43	$\pm$	0.09	&	0.87	$\pm$	0.11	&	0.88	$\pm$	0.11	&	0.27	$\pm$	0.10	&	0.32	$\pm$	0.11	&	-0.52	$\pm$	0.13	&	2.99	$\pm$	0.34	&	I,II,I+II	&	/	&	0;0;0	\\
11	&	XMMU J183458.3-330548	&	1.17	&	0.58	$\pm$	0.12	&	0.83	$\pm$	0.13	&	0.35	$\pm$	0.10	& \textit{0.02	$\pm$	0.07}	&	0.17	$\pm$	0.13	&	-0.88$^{+0.35}_{-0.12}$	&	2.35	$\pm$	0.35	&	II,I+II		&	/	&	0;1;0	\\
12	&	XMMU J183509.6-325607	&	1.19	&	0.43	$\pm$	0.10	&	0.43	$\pm$	0.09	&	0.62	$\pm$	0.10	& \textit{0.02	$\pm$	0.05}	&	-0.04	$\pm$	0.15	&	-0.91$^{+0.16}_{-0.09}$	&	1.74	$\pm$	0.26	&	I,I+II		&	V	&	0;0;0	\\
13	&	XMMU J183536.7-325217	&	1.19	&	0.51	$\pm$	0.09	&	0.75	$\pm$	0.10	&	0.35	$\pm$	0.08	&	0.14	$\pm$	0.08	&	0.19	$\pm$	0.11	&	-0.43	$\pm$	0.24	&	2.35	$\pm$	0.30	&	I,II,I+II	&	/	&	0;0;0	\\
14	&	XMMU J183507.4-325911	&	1.18	&	0.16	$\pm$	0.06	&	0.54	$\pm$	0.08	&	0.51	$\pm$	0.08	& \textit{0.01	$\pm$	0.03}	&	0.55	$\pm$	0.14	&	-0.98$^{+0.11}_{-0.02}$	&	1.45	$\pm$	0.24	&	I,II,I+II	&	/	&	0;0;0	\\
15	&	XMMU J183526.9-325422	&	1.27	&	0.16	$\pm$	0.12	&	0.37	$\pm$	0.17	&	1.31	$\pm$	0.25	&	0.94	$\pm$	0.25	&	0.35	$\pm$	0.37	&	-0.17	$\pm$	0.16	&	3.46	$\pm$	0.68	&	I,I+II		&	/	&	0;0;0	\\
16	&	XMMU J183638.7-325144	&	1.27	&	1.49	$\pm$	0.30	&	1.57	$\pm$	0.30	&	0.98	$\pm$	0.28	& \textit{0.11	$\pm$	0.26}	&	0.08	$\pm$	0.13	&	-0.45	$\pm$	0.33	&	5.88	$\pm$	1.05	&	I,II,I+II	&	/	&	0;0;0	\\
17	&	XMMU J183526.1-330933	&	1.34	&	0.53	$\pm$	0.16	&	1.01	$\pm$	0.20	&	1.20	$\pm$	0.21	& \textit{0.13	$\pm$	0.13}	&	0.31	$\pm$	0.16	&	-0.81	$\pm$	0.17	&	3.11	$\pm$	0.49	&	I,I+II		&	V	&	0;1;0	\\
18	&	XMMU J183455.9-324948	&	1.31	&	0.79	$\pm$	0.17	&	0.64	$\pm$	0.16	&	0.66	$\pm$	0.15	&	0.20	$\pm$	0.18	&	-0.11	$\pm$	0.16	&	-0.53	$\pm$	0.33	&	2.79	$\pm$	0.46	&	I,I+II		&	/	&	0;0;0	\\
19	&	XMMU J183629.8-330208	&	1.28	&	0.46	$\pm$	0.12	&	0.43	$\pm$	0.11	&	0.66	$\pm$	0.13	&	0.30	$\pm$	0.14	&	-0.05	$\pm$	0.17	&	-0.34	$\pm$	0.21	&	1.88	$\pm$	0.33	&	II,I+II		&	/	&	0;0;0	\\
20	&	XMMU J183542.4-331107	&	1.29	&	1.14	$\pm$	0.22	&	1.07	$\pm$	0.20	&	0.31	$\pm$	0.14	&	0.61	$\pm$	0.24	&	0.00	$\pm$	0.13	&	0.28	$\pm$	0.26	&	3.02	$\pm$	0.51	&	I,I+II		&	/	&	1;0;0	\\
21	&	XMMU J183459.5-325430	&	1.46	&	0.30	$\pm$	0.12	&	0.52	$\pm$	0.13	&	0.42	$\pm$	0.13	&	0.23	$\pm$	0.15	&	0.26	$\pm$	0.22	&	-0.28	$\pm$	0.34	&	1.73	$\pm$	0.30	&	I+II		&	/	&	0;0;0	\\
22	&	XMMU J183632.8-330509	&	1.20	&	0.81	$\pm$	0.16	&	0.84	$\pm$	0.16	&	0.73	$\pm$	0.15	& \textit{0.06	$\pm$	0.11}	&	0.06	$\pm$	0.14	&	-0.76$^{+0.26}_{-0.24}$	&	2.61	$\pm$	0.43	&	I,I+II		&	V	&	1;0;1	\\
23	&	XMMU J183539.1-324747	&	1.90	&	0.82	$\pm$	0.20	&	0.66	$\pm$	0.18	& \textit{0.00	$\pm$	0.09}	& \textit{0.03	$\pm$	0.11}	&	-0.11	$\pm$	0.18	&	-			&	2.15	$\pm$	0.40	&	I+II		&	/	&	3;0;1	\\
24	&	XMMU J183450.7-325705	&	1.52	&	0.27	$\pm$	0.12	&	0.57	$\pm$	0.15	&	0.33	$\pm$	0.13	& \textit{0.13	$\pm$	0.15}	&	0.35	$\pm$	0.23	&	-0.44	$\pm$	0.48	&	1.57	$\pm$	0.31	&	I+II		&	/	&	0;2;0	\\
25	&	XMMU J183514.8-325354	&	1.27	&	0.36	$\pm$	0.10	&	0.47	$\pm$	0.11	&	0.62	$\pm$	0.12	&	0.14	$\pm$	0.11	&	0.11	$\pm$	0.18	&	-0.64	$\pm$	0.24	&	1.90	$\pm$	0.33	&	II,I+II		&	/	&	0;0;0	\\
26	&	XMMU J183453.2-325851	&	1.27	&	0.75	$\pm$	0.12	&	0.40	$\pm$	0.10	& \textit{0.00	$\pm$	0.02}	& \textit{0.03	$\pm$	0.07}	&	-0.29	$\pm$	0.13	&	-			&	1.29	$\pm$	0.28	&	II,I+II		&	/	&	0;0;0	\\
27	&	XMMU J183611.0-330725	&	1.53	& \textit{0.00	$\pm$	0.04}	& \textit{0.11	$\pm$	0.11}	&	0.67	$\pm$	0.17	&	0.55	$\pm$	0.21	&	1.00$^{+0.00}_{-0.70}$	&	-0.10	$\pm$	0.23	&	1.34	$\pm$	0.32	&	I+II		&	/	&	0;0;0	\\
28	&	XMMU J183512.1-330324	&	1.18	&	0.74	$\pm$	0.10	&	0.43	$\pm$	0.08	&	0.11	$\pm$	0.05	&	0.11	$\pm$	0.06	&	-0.26	$\pm$	0.11	&	-0.03	$\pm$	0.37	&	1.70	$\pm$	0.26	&	I,II,I+II	&	/	&	0;1;1	\\
29	&	XMMU J183523.7-324836	&	1.68	&	0.91	$\pm$	0.20	&	0.35	$\pm$	0.13	& \textit{0.02	$\pm$	0.06}	& \textit{0.00	$\pm$	0.11}	&	-0.44	$\pm$	0.18	&	-			&	1.82	$\pm$	0.35	&	I+II		&	/	&	0;1;0	\\
30	&	XMMU J183448.4-330107	&	1.79	&	0.91	$\pm$	0.19	&	0.48	$\pm$	0.15	& \textit{0.00	$\pm$	0.04}	& \textit{0.09	$\pm$	0.14}	&	-0.30	$\pm$	0.17	&	-			&	1.68	$\pm$	0.34	&	I+II		&	/	&	0;0;0	\\
31	&	XMMU J183516.1-325724	&	1.31	&	0.29	$\pm$	0.09	&	0.44	$\pm$	0.10	&	0.34	$\pm$	0.09	& \textit{0.02	$\pm$	0.05}	&	0.21	$\pm$	0.19	&	-0.87$^{+0.24}_{-0.13}$	&	1.46	$\pm$	0.30	&	II,I+II		&	/	&	0;2;0	\\
32	&	XMMU J183514.1-325623	&	1.90	&	0.43	$\pm$	0.11	&	0.32	$\pm$	0.10	&	0.16	$\pm$	0.08	& \textit{0.00	$\pm$	0.05}	&	-0.15	$\pm$	0.20	&	-1.00$^{+0.58}_{-0.00}$	&	1.15	$\pm$	0.23	&	I+II		&	/	&	1;1;1	\\
33	&	XMMU J183541.1-325339	&	1.39	&	0.18	$\pm$	0.08	&	0.58	$\pm$	0.11	&	0.35	$\pm$	0.09	&	0.08	$\pm$	0.06	&	0.53	$\pm$	0.16	&	-0.55	$\pm$	0.23	&	1.22	$\pm$	0.24	&	I,I+II		&	V	&	1;0;0	\\
34	&	XMMU J183622.4-325338	&	1.69	&	0.22	$\pm$	0.12	&	0.63	$\pm$	0.17	&	0.38	$\pm$	0.14	& \textit{0.00	$\pm$	0.07}	&	0.48	$\pm$	0.23	&	-1.00$^{+0.38}_{-0.00}$	&	1.60	$\pm$	0.34	&	I+II		&	/	&	0;0;0	\\
35	&	XMMU J183630.2-330708	&	1.32	&	0.40	$\pm$	0.13	&	1.09	$\pm$	0.18	&	0.43	$\pm$	0.13	& \textit{0.01	$\pm$	0.06}	&	0.46	$\pm$	0.15	&	-0.93$^{+0.26}_{-0.07}$	&	2.32	$\pm$	0.41	&	I,I+II		&	/	&	0;0;0	\\ \hline
\end{tabular}
\end{small}
\end{center}
\end{table}

\clearpage
\addtocounter{table}{-1}

\begin{table}[htbp]
\begin{small}
\begin{center}
\begin{tabular}[c]{|c|c|c|c|c|c|c|c|c|c|c|c|c|}
\hline
(1)	& 	(2)			& 	(3)	& 	(4)			& 	(5)			& 	(6)			& 	(7)			& 	(8)			& 	(9)			& 	(10)			& 	(11)	& 	(12)	&	(13)	\\
SRC 	&	SOURCE NAME		&	ERR	& 	RATE 1			&	RATE 2			&	RATE 3			& 	RATE 4			& 	HR1			&	HR2			& 	FLUX			&	OBS	&	VAR	&	CTP	\\
ID	&				&	($''$)	& ($10^{-3}$c s$^{-1}$)		& ($10^{-3}$c s$^{-1}$)		& ($10^{-3}$c s$^{-1}$)		& ($10^{-3}$c s$^{-1}$)		& 				& 				& 	($10^{-14}$ cgs)	&		&		&		\\ \hline
36	&	XMMU J183515.1-325016	&	1.83	&	0.97	$\pm$	0.19	& \textit{0.00	$\pm$	0.10}	& \textit{0.00	$\pm$	0.04}	& \textit{0.02	$\pm$	0.10}	&	-1.00$^{+0.20}_{-0.00}$	&	-			&	1.39	$\pm$	0.33	&	I+II	&	/	&	2;5;1	\\
37	&	XMMU J183628.7-330409	&	1.36	&	0.60	$\pm$	0.13	&	0.65	$\pm$	0.13	&	0.30	$\pm$	0.10	&	0.19	$\pm$	0.15	&	0.04	$\pm$	0.14	&	-0.04	$\pm$	0.38	&	1.75	$\pm$	0.34	&	II,I+II	&	V	&	0;0;0	\\
38	&	XMMU J183442.8-325840	&	2.01	&	0.26	$\pm$	0.14	&	0.62	$\pm$	0.17	&	0.46	$\pm$	0.16	& \textit{0.06	$\pm$	0.13}	&	0.40	$\pm$	0.25	&	-0.75$^{+0.45}_{-0.25}$	&	1.74	$\pm$	0.37	&	I+II	&	/	&	1;2;1	\\
39	&	XMMU J183614.3-325949	&	1.71	&	0.10	$\pm$	0.08	&	0.40	$\pm$	0.11	&	0.21	$\pm$	0.09	&	0.12	$\pm$	0.10	&	0.61	$\pm$	0.28	&	-0.26	$\pm$	0.44	&	1.11	$\pm$	0.23	&	I+II	&	/	&	1;0;0	\\
40	&	XMMU J183538.4-324701	&	1.99	&	0.35	$\pm$	0.18	&	0.60	$\pm$	0.17	&	0.23	$\pm$	0.14	&	0.28	$\pm$	0.24	&	0.27	$\pm$	0.28	&	0.11	$\pm$	0.52	&	2.17	$\pm$	0.44	&	I+II	&	/	&	0;0;0	\\
41	&	XMMU J183548.1-331137	&	1.78	& \textit{0.17	$\pm$	0.20}	&	1.06	$\pm$	0.32	&	0.92	$\pm$	0.33	&	0.69	$\pm$	0.40	&	0.73$^{+0.27}_{-0.29}$	&	-0.15	$\pm$	0.33	&	3.04	$\pm$	0.69	&	I+II	&	/	&	0;0;0	\\
42	&	XMMU J183538.5-325139	&	1.74	&	0.82	$\pm$	0.23	& \textit{0.07	$\pm$	0.11}	&	0.29	$\pm$	0.15	& \textit{0.13	$\pm$	0.15}	&	-0.83$^{+0.23}_{-0.17}$	&	-0.39	$\pm$	0.56	&	1.89	$\pm$	0.42	&	I	&	V	&	0;0;1	\\
43	&	XMMU J183512.8-324755	&	1.66	&	1.83	$\pm$	0.50	& \textit{0.15	$\pm$	0.23}	& \textit{0.19	$\pm$	0.26}	& \textit{0.00	$\pm$	0.19}	&	-0.85$^{+0.22}_{-0.15}$	&	-			&	3.81	$\pm$	0.93	&	I	&	V	&	0;0;0	\\
44	&	XMMU J183529.7-324734	&	2.34	&	1.19	$\pm$	0.36	& \textit{0.18	$\pm$	0.19}	& \textit{0.08	$\pm$	0.15}	& \textit{0.04	$\pm$	0.19}	&	-0.74	$\pm$	0.24	&	-			&	2.78	$\pm$	0.66	&	I	&	V	&	3;3;1	\\
45	&	XMMU J183632.9-325653	&	1.61	&	0.71	$\pm$	0.22	&	0.76	$\pm$	0.24	& \textit{0.15	$\pm$	0.16}	& \textit{0.00	$\pm$	0.07}	&	0.04	$\pm$	0.22	&	-			&	2.13	$\pm$	0.49	&	II	&	V	&	0;0;0	\\
46	&	XMMU J183515.9-324703	&	2.11	&	1.31	$\pm$	0.49	&	0.61	$\pm$	0.37	& \textit{0.06	$\pm$	0.18}	&	1.24	$\pm$	0.66	&	-0.36	$\pm$	0.31	&	0.91$^{+0.09}_{-0.27}$	&	4.74	$\pm$	1.10	&	II	&	/	&	0;1;0	\\ \hline
\end{tabular}
\end{center}
\
Key to Table: Col.(1) = Source ID number; Col.(2) = source XMM name; Col.(3) = position error; Col.(4),(5),(6),(7) = MOS1+MOS2 count rates in the energy ranges 0.5--1, 1--2, 2--5, and 4 = 5--10~keV, respectively (rates consistent with false detections are reported in italic); Col.(8),(9) = count rate hardness ratios, defined as HR1 = (2-1)/(2+1) and HR2 = (4-3)/(4+3) (values are omitted when in both energy ranges the measured count rates are consistent with a false detection); Col.(10) = flux in the 0.5--10~keV energy range, in units of 10$^{-14}$~erg~cm$^{-2}$~s$^{-1}$, calculated as reported in the text; Col.(11) = XMM observation with a positive source detection; Col.(12) = flag indicating a source variability; Col. (13) = number of GSC, USNO--NOMAD and 2MASS counterparts, respectively, found considering a circular uncertainty region with a radius of three times the source position error.
\end{small}
\label{sources}
\end{table}

\end{landscape}

\end{document}